\documentclass[11pt, a4]{article}

\usepackage{amsmath}
\usepackage{amsfonts}
\usepackage{amssymb}
\usepackage{mathrsfs}
\usepackage[T1]{fontenc}
\usepackage[latin1]{inputenc}
\usepackage{textcomp}
\usepackage[english]{babel}
\usepackage{epsfig}
\usepackage{verbatim}
\usepackage{graphicx}
\usepackage{epstopdf}

\numberwithin{equation}{section} 
\providecommand{\abs}[1]{\lvert#1\rvert}

\def\a{\alpha}
\def\b{\beta}

\def\r{\rho}

\newcommand{\aq}{\alpha^{2}}
\newcommand{\bq}{\beta^{2}}
\newcommand{\ha}[1]{h^{#1}}
\newcommand{\va}[1]{v^{#1}}

\newcommand{\p}{\prime}

\numberwithin{equation}{section}

\def\mK{{\mathcal K}}
\def\tr{{\rm tr}\,}

\def\be{\begin{equation}}
\def\ee{\end{equation}}
\def\bea{\begin{eqnarray}}
\def\eea{\end{eqnarray}}

\begin{document}

\date{\mbox{}}

\title{
\vspace{-2.0cm}
\vspace{2.0cm}
{\bf \huge Characterising Vainshtein Solutions in Massive Gravity}
 \\[8mm]
}

\author{
Fulvio Sbis\`a$^{1,2}$\thanks{fulvio.sbisa@port.ac.uk}, Gustavo Niz$^{1,3}$\thanks{g.niz@ugto.mx}, Kazuya Koyama$^1$\thanks{kazuya.koyama@port.ac.uk}, Gianmassimo Tasinato$^1$\thanks{gianmassimo.tasinato@port.ac.uk}
%.
\\[8mm]
\normalsize\it
$^1$ Institute of Cosmology \& Gravitation, University of Portsmouth,\\
\normalsize\it Dennis Sciama Building, Portsmouth, PO1 3FX, United Kingdom\vspace{.5cm} \\
\normalsize\it
$^2\,$ Dipartimento di Fisica dell'Universit\`a di Milano\\
       {\it Via Celoria 16, I-20133 Milano} \\
       and \\
\normalsize\it
       INFN, Sezione di Milano, \\
       {\it Via Celoria 16, I-20133 Milano}\vspace{.5cm} \\
\normalsize\it
$^3$ Departamento de F\'{\i}sica, Universidad de Guanajuato,\\
{\it DCI, Campus Le\' on, C.P. 37150, Le\' on, Guanajuato, M\' exico.}\\[.3em]
}

\maketitle

\setcounter{page}{1}
\thispagestyle{empty}

\begin{abstract}

\noindent
We study static, spherically symmetric solutions in a recently proposed ghost-free model of non-linear massive gravity. We focus on a branch of solutions where the helicity-0 mode can be strongly coupled within certain radial regions, giving rise to the Vainshtein ef\mbox{}fect. We truncate the analysis to scales below the gravitational Compton wavelength, and consider the weak f\mbox{}ield limit for the gravitational potentials, while keeping all non-linearities of the helicity-0 mode. We determine analytically the number and properties of local solutions which exist asymptotically on large scales, and of local (inner) solutions which exist on small scales. We f\mbox{}ind two kinds of asymptotic solutions, one of which is asymptotically f\mbox{}lat, while the other one is not, and also two types of inner solutions, one of which displays the Vainshtein mechanism, while the other exhibits a self-shielding behaviour of the gravitational f\mbox{}ield. We analyse in detail in which cases the solutions match in an intermediate region. The asymptotically f\mbox{}lat solutions connect only to inner conf\mbox{}igurations displaying the Vainshtein mechanism, while the non asymptotically f\mbox{}lat solutions can connect with both kinds of inner solutions. We show furthermore that there are some regions in the parameter space where global solutions do not exist, and characterise precisely in which regions of the phase space the Vainshtein mechanism takes place.
\end{abstract}

\def\eq#1{eq.~(\ref{#1})}

\smallskip

\section{Introduction}

Promoting Einstein's gravity to a classical theory of a massive graviton is a theoretical challenge.
The appearance of a ghost instability and the fact that in the massless limit solutions may not reduce to
those of General Relativity (GR) have been the two main obstacles for a successful Lagrangian construction
\cite{Fierz:1939ix}. Both of these problems are related to the scalar sector of the theory.
In recent years, new developments have opened a window into a ghost-free model for massive gravity
\cite{FPextension, resummation, rachel2}, based on the request that the equation of motion for the scalar mode does not contain derivatives of order higher than two \cite{nicolis}.
%Galilean symmetry for the scalar mode \cite{nicolis}.
However, with respect to the second issue the story is not entirely settled, as we will discuss here.

In this massive gravity theory, spherically symmetric solutions do not obey a uniqueness principle, like Birkhof\mbox{}f's theorem in General Relativity. As a result, the theory exhibits two classes of static spherically symmetric solutions, which dif\mbox{}fer in how the scalar sector couples to graviton. The f\mbox{}irst class of solutions naturally contains (Anti) de Sitter-Schwarzschild solutions, where the cosmological constant is proportional to the graviton's mass \cite{us1,us2,niew}. Some of these solutions are very similar to those found many years ago in the simple Fierz-Pauli model \cite{salam}, and what characterises them is that the scalar mode is strongly coupled everywhere, which avoids a discontinuous connection to GR in
the massless limit. Therefore, at the level of the background, these solutions are indistinguishable
from their counterparts in General Relativity, but do present dif\mbox{}ferences at f\mbox{}irst orders in
perturbations. Interestingly, in some of these solutions, scalar and vector perturbations are not
dynamical at f\mbox{}irst order in perturbations, and only the tensor modes dif\mbox{}fer signif\mbox{}icantly from General Relativity \cite{shinji,new,new2}. If the scalar and vector modes remain non-dynamical at higher orders in perturbation theory, then only the detection of gravitational waves can rule out these solutions in favour of $\Lambda$CDM.

In the second class of solutions the story is not that clear. The scalar sector is not strongly coupled
everywhere, and the solutions behave very dif\mbox{}ferently as a function of the radial coordinate.
The purpose of this paper is to exhibit the rich  structure of such class of solutions, and show
whether one can recover GR solutions in the massless limit, at least within some radial region. This
phenomenon of recovering GR predictions within some macroscopic radius from a mass source is known as
the Vainshtein mechanism \cite{vainshtein}. In the range where the solution reduces to GR the scalar
f\mbox{}ield becomes strongly coupled. Therefore, it is imperative to include non-linearities in the scalar
sector in order to study the Vainshtein mechanism. Solutions presenting these behaviour in massive
gravity where found in \cite{Damour:2002gp}, and more recently in the ghost-free model considered
here in \cite{us1, us2, david, sjors}. Unfortunately, these solutions do not cover the whole parameter
space of the theory as we will describe here.

We organise the paper as follows: in Section 2 we give an overview of the main equations describing
the theory and spherically symmetric solutions, with particular attention to this second branch. In
Section 3 we present an exhaustive analysis of the vacuum equations, with particular attention to
the behaviour of their solutions near the origin and towards inf\mbox{}inity. In Section 4 we present our
main results, including the parameter space in which GR solutions are recovered via the Vainshtein
mechanism and numerical solutions of the vacuum equations for some representative cases. Finally,
we present some conclusions in Section 5.

\section{Ghost-free massive gravity and spherically symmetric ans\"atze}

We consider the following Lagrangian for massive gravity, a non-linear
extension of Fierz-Pauli theory proposed in \cite{resummation}
\be\label{genlag}
{\cal L} = \frac{M_{Pl}^2}{2}\,\sqrt{-g}\left( R - {\cal U}\right).
\ee
The potential ${\cal U}$ depends on a dimension-full parameter $m$, which sets the graviton
mass scale, and on two dimensionless parameters $\alpha_3$ and $\alpha_4$. It has the following
functional form
\be
{\cal U}= -m^2\left[{\cal U}_2+\alpha_3\, {\cal U}_3+\alpha_4\, {\cal U}_4\right],
\label{potentialU}
\ee
with
\bea
{\cal U}_2&=&(\tr\mK)^2-\tr (\mK^2),\nonumber \\
{\cal U}_3&=&(\tr \mK)^3 - 3 (\tr \mK)(\tr \mK^2) + 2 \tr \mK^3,\nonumber \\
{\cal U}_4 &=& (\tr \mK)^4 - 6 (\tr \mK)^2 (\tr \mK^2)
+ 8 (\tr \mK)(\tr \mK^3) + 3 (\tr \mK^2)^2 - 6 \tr \mK^4 .\nonumber
\eea
The tensor ${\cal K}_{\mu}^{\ \nu}$ is def\mbox{}ined as \cite{resummation}
\bea
{\mathcal K}_{\mu}^{\ \nu} &\equiv &\delta_{\mu}^{\ \nu}-\left(\sqrt{g^{-1} \Sigma}\right)_{\mu}^{\ \nu}\,\,,
\eea
where the square root of a tensor is formally understood as
$\sqrt{{\cal M}}_{\mu}^{\ \alpha}\sqrt{\cal M}_{\alpha}^{\ \nu}={\cal M}_{\mu}^{\ \nu}$,
for any tensor ${\cal M}_\mu^{\,\,\nu}$. The tensor $\Sigma_{\mu\nu}$ is a f\mbox{}iducial metric, which
also makes the theory reparametrisation invariant by means of four scalars $\phi^\mu$, and it is given by
\be
\Sigma_{\mu\nu}=\partial_\mu\phi^a\partial_\nu\phi^b\eta_{ab}.
\ee
The theory def\mbox{}ined by (\ref{genlag}) has Minkowski spacetime as solution, hence one can rewrite the metric
$g_{\mu\nu}$ and the scalars $\phi^\mu$ as deviations from f\mbox{}lat space, namely
\be
g_{\mu\nu}=\eta_{\mu\nu}+h_{\mu\nu},\qquad \qquad \phi^\mu=x^\mu+\pi^\mu,
\ee
where $x^\mu$ are the usual cartesian coordinates spanning $\eta_{\mu\nu}$.  Therefore, a change
of coordinates $x^{\mu} \to x^{\mu} + \xi^{\mu}$ should be accompanied by the following
transformation of the St\"uckelberg f\mbox{}ield $\pi^\mu$,
\be\label{pi_trans}
\pi^{\mu} \to \pi^{\mu} + \xi^{\mu},
\ee
in order to recover full dif\mbox{}feomorphism invariance. Moreover, we choose the {\it unitary}
gauge, where $\pi^\mu=0$, and the potential (\ref{potentialU}) considerably simplif\mbox{}ies.

Since we are intested in spherically symmetric solutions, we use the most general static ans\"atze, given by
\be\label{genmetr}
d s^2\,=\,-C(r) \,d t^2+A(r)\, d r^2 +2 D(r)\, dt dr+B(r) d \Omega^2,
\ee
where $d \Omega^2 = d \theta^2 + \sin^2 \theta d \phi^2$.
We choose to write the non-dynamical f\mbox{}lat metric as $ds^2 = -dt^2 + dr^2 + r^2 d \Omega^2$.
It should be noticed that this is not a coordinate choice, but a way to simplify the expressions.
We plug the previous metric into the Einstein equations $G_{\mu \nu}=T^{{\cal U}}_{\mu \nu}$,
where the energy momentum tensor is def\mbox{}ined as
$T^{{\cal U}}_{\mu \nu}\,=\frac{m^2}{\sqrt{-g}}\,\frac{ \delta \sqrt{-g}\ {\cal U}}{\delta g^{\mu \nu}}$.
The Einstein tensor $G_{\mu\nu}$ satisf\mbox{}ies the identity $D(r)\, G_{tt}+C(r)\,G_{tr}\,=\,0$, which
implies the algebraic constraint $0= D(r)\, T^{{\cal U}}_{tt}+C(r)\,T^{{\cal U}}_{tr}$. This last
equation reduces to $D(r)\left(b_0 r-\sqrt{B(r)}\right)$, where $b_0$ is a function of $\alpha_3$ and $\alpha_4$ only \cite{new}. This constraint is solved in two possible ways, def\mbox{}ining two class of solutions: either the metric is diagonal $D=0$, or $B=b_0^2 r^2$. Notice that this classif\mbox{}ication only holds in the unitary gauge, since one can always map the metric from one class to the other by a coordinate transformation, but to the price of exciting components of $\pi^\mu$.

The class of solutions with a non-diagonal metric leads to Schwarzschild or Schwarzschild-de Sitter
solutions \cite{us1, us2, niew}, as explained in the introduction. In this sector of the
theory GR is recovered by means of a strongly coupled $\pi^\mu$ everywhere \cite{shinji, new}.
However, in the other class of solutions, where the metric is diagonal in the unitary gauge,
the situation is dif\mbox{}ferent. As we discuss in the next section, $\pi^\mu$ may or may not be strongly coupled: it could be strongly coupled within certain radial region, leading to a Vainshtein ef\mbox{}fect. This branch is the one that concerns us here, so from now on we will only consider static diagonal spherically symmetric solutions in the unitary gauge (see \cite{Deffayet:2011rh} for a similar discussion on bimetric solutions in the unitary gauge).

The problem of f\mbox{}inding exact vacuum solutions in this branch is an open question, but one
can make progresses by considering perturbations (not necessarily small) from f\mbox{}lat space, and
the following ansatz results adequate for this purpose,
\be
ds^2 = - \Big(1+N(r)\Big)^2 dt^2 + \Big(1+F(r)\Big)^{-1} dr^2 + r^2 \Big(1+H(r)\Big)^{-2} d \Omega^2.
\label{diagonal}
\ee
In order to analyse the system, it is convenient to introduce a new radial coordinate
\be
\rho = \frac{r}{1+H(r)}\,,
\label{chcoord}
\ee
so that the linearised metric is expressed as
\be
ds^2 = - (1 + n) dt^2 + (1 - f) d\rho^2
+ \rho^2 d \Omega^2,
\ee
where $f(\rho) =  F\big(r(\rho)\big) - 2  h(\rho) - 2 \rho  h'(\rho)$, $n(\rho)=2N\big(r(\rho)\big)$, $h(\rho)=H\big(r(\rho)\big)$ and a prime denotes a derivative with respect to $\rho$.
As discussed above, one should be careful with this change of coordinates, since, after f\mbox{}ixing a
gauge, a change of frame in the metric modif\mbox{}ies the St\"uckelberg f\mbox{}ield $\pi^\mu$ as well. It
turns out that this coordinate transformation excites the radial component of $\pi^\mu$, which
explicitly is $\pi^\rho=\rho h$. Therefore, from now on one can think of $h$ as simply being the
only non-zero component of the St\"uckelberg f\mbox{}ield $\pi^\mu$. At linear order, the equations for
the functions $n(\rho)$, $f(\rho)$ and $h(\rho)$ in the new variable $\rho$ are
\bea\label{Neq}
0&=& \left(m^2\rho^2+2\right)f+2 \rho \left(f'+m^2 \rho^2
h'+3 \, m^2 \rho h \right), \\
0&=&  \frac{1}{2}m^2 \rho^2 (n-4 h) -\rho \, n'-f, \label{feq}
\\
0&=&  f +\frac{1}{2}\rho \, n'\label{const}.
\eea
In this linear expansion, the solutions for $n$ and $f$ are
\bea
n &=& - \frac{8 G M}{3 \rho} e^{- m \rho} \label{linsoln}\\
f &=& -\frac{4 G M}{3 \rho} (1 + m \rho) e^{- m \rho} \label{linsolf}
\eea
where we f\mbox{}ix the integration constant so that $M$ is the mass of a point particle at the origin,
and $8 \pi G = M_{pl}^{-2}$. These solutions exhibit the vDVZ discontinuity, since the post-Newtonian
parameter $\gamma=f/n$ is $\gamma=\frac{1}{2}(1+m\rho)$, which in the massless limit reduces to $\gamma=1/2$,
in disagreement with GR and Solar system observations.

However, in order to understand what really happens in this limit, we must also analyse the behaviour
of $h$, or equivalently $\pi^\rho$, as $m\rightarrow 0$. To do this, we consider scales below the
Compton wavelength $m \rho \ll 1$, and at the same time ignore higher order terms in $G M$. Under
these approximations, the equations of motion can still be truncated to linear order in $f$
and $n$, but since $h$ is not necessarily small, we have to keep all non-linear terms in $h$. In
other words, we take the usual weak f\mbox{}ield limit for the metric f\mbox{}ields, but keep all non-linearities
in the St\"uckelberg f\mbox{}ield, since we expect regions where this f\mbox{}ield is strongly coupled. As shown
in \cite{us2}, the f\mbox{}ield equations reduce to the following system of coupled equations for the
f\mbox{}ields $f$, $n$, $h$:
\begin{gather}
\label{solf}
f = - 2 \frac{G M}{\rho} - (m \rho)^2 \left[
h - (1+ 3\alpha_3)h^2+(\alpha_3+4\alpha_4)h^3\right] \\[2mm]
\label{soln}
\rho\,  n' = \frac{2 G M}{\rho} - (m \rho)^2 \left[
h - (\alpha_3 +4\alpha_4) h^3\right] \\[2mm]
\begin{split}
 \frac{G M}{\rho} \left[1 - 3(\alpha_3+4\alpha_4)h^2\right] &= - (m \rho)^2
\Big\{\frac{3}{2} h -  3 (1 + 3\alpha_3)h^2  +
\left[(1 + 3\alpha_3)^2 + 2(\alpha_3 +4\alpha_4)\right]h^3 \Big. \\
& \hspace{2.1cm} \Big. - \frac{3}{2}(\alpha_3 +4\alpha_4)^2 h^5 \Big\} \label{solh}
\end{split}
\end{gather}
These equations can also be obtained directly from the decoupling theory \cite{FPextension,deRham:2010tw}, as it
was shown in \cite{us2}. The previous expressions are the starting point of our analysis and
classif\mbox{}ications of solutions, which we will discuss in the next section.

\section{Classifying Solutions}

Our aim is to scan the $(\a_3, \a_4)$ parameter phase space of theories, to understand how many solutions the system (\ref{solf})-(\ref{solh}) admits, and characterise their (asymptotic) geometrical properties.

Since the last equation (\ref{solh}) does not contain $f$ and $n$, the f\mbox{}ield $h$ obeys a decoupled equation, thus once a solution of this equation is found, the gravitational potentials $f$, $n$ are uniquely determined (up to an integration constant) by the other two equations (\ref{solf}) and (\ref{soln}). Therefore, we f\mbox{}irst focus on classifying the number and properties of solutions of (\ref{solh}) in every point of the phase space, and then discuss the behaviour of the gravitational potentials correspondent to these solutions.

\subsection{The quintic equation}

For notational convenience, we def\mbox{}ine $\a \equiv 1 + 3 \, \a_3$ and $\b \equiv \a_3 + 4 \, \a_4 \,$. Therefore, the system that determines the gravitational potentials in terms of $h$ takes the form
\begin{align}
\label{solfab}
f &= - 2 \, \frac{G M}{\rho} - (m \rho)^2 \Big(
h - \alpha h^2 + \beta h^3 \Big) \\[2mm]
\label{solnab}
n' &= 2 \, \frac{G M}{\rho^2} - m^2 \rho \, \Big( h - \b h^3 \Big)
\end{align}

\noindent while the equation which determines $h$ reduces to
\be
\label{quintic}
\frac{3}{2} \, \bq \, \ha{5}(\r) - \Big( \aq + 2 \b \Big) \, \ha{3}(\r) + 3 \, \Big( \a + \b A(\r) \Big) \, \ha{2}(\r)
- \frac{3}{2} \, h(\r) - A(\r) = 0
\ee
where $A(\r) = \big( \r_{v} / \r \big)^{3}$ and $\r_{v}$ is the Vainshtein radius def\mbox{}ined
as $\r_{v} \equiv \big( G M / m^{2} \big)^{\! 1/3}$. The last equation is an algebraic equation in
$h$, $A$, $\a$ and $\b$; at f\mbox{}ixed  $\r$, $\a$ and $\b$ it is, in fact, a polynomial equation of f\mbox{}ifth degree in $h$, except for the special case $\b = 0$. In this particular case of $\beta=0$, the equation for $h$ becomes a cubic equation and it is possible to obtain solutions for $h$ and the metric perturbations exactly. These solutions were studied in \cite{us1,us2} and it was shown that the solutions exhibit the Vainshtein mechanism. Therefore, in what follows, we assume $\beta \neq 0$. It is dif\mbox{}f\mbox{}icult to f\mbox{}ind analytical solutions to this quintic equation, and it is not easy even to understand how many solutions it admits. In fact, even if a local solution is found in a radial interval, it is not, in general, possible to extend it to the whole radial domain, as we will explicitly see in the next section.

Nevertheless, it is possible to determine exactly how many local solutions exist in a neighbourhood of
$\r = +\infty$, which we refer as {\it asymptotic solutions}, and also how many local solutions exist in a
neighbourhood of $\r = 0^{+}$, which we call {\it inner solutions}. Furthermore, we can f\mbox{}ind analytically
their leading behaviour as a function of $\rho$. Any global solution of (\ref{quintic}) should necessarily interpolate between one of the asymptotic solutions and one of the inner solutions. Therefore, our aim is to understand, for each point in the  $(\a, \b)$ phase space, whether and how the above solutions match.

A systematic approach to Vainshtein ef\mbox{}fects in covariant Galileon theory was performed in  \cite{Kaloper:2011qc}, and in general scalar-tensor theories  \cite{Kimura:2011dc}. However, in massive gravity a systematic treatment was not fully performed. The main dif\mbox{}ference between massive gravity and standard Galileon theories is that in the latter case the f\mbox{}ifth power of $h$ in equation (\ref{solh}) is absent. The reason of this becomes particularly transparent when focussing on the decoupling limit. Then one can see that, for $\beta\neq 0$, there is a  mixing between $h_{\mu\nu}$ and a combination of derivatives of the scalar graviton polarization, that cannot be removed by f\mbox{}ield redef\mbox{}initions. Such derivative mixing of the scalar with gravity is not included in standard Galileon models, and the corresponding equations of motion  consequently miss its ef\mbox{}fect. Implementing a suitable coordinate transformation \cite{us2}, one can recognize that  precisely the aforementioned mixing gives rise to the quintic term in $h$ in the equation (\ref{solh}). As we will discuss in what follows, the analysis associated with the quintic equation turning on a $\beta\neq 0$ enriches considerably the properties of the phase space with respect to the case $\beta=0$.

\subsection{Phase space analysis}

To be able to describe how the matching works in all the phase space, in principle we should study
separately every point ($\a$, $\b$). However, this is not necessary since equation (\ref{quintic}) obeys a
remarkable symmetry: def\mbox{}ining the quintic function as
\be
\label{quinticFunction}
q \, \big( h; A, \a, \b \big) \equiv \frac{3}{2} \, \bq \, \ha{5} - \big( \aq + 2 \b \big) \, \ha{3} + 3 \, \big( \a + \b A \big) \, \ha{2}
- \frac{3}{2} \, h - A
\ee
it is simple to see that
\be
\label{symmetry}
q \, \Big( \frac{h}{k} ; \frac{A}{k} , k \, \a, k^2 \b \Big) = \frac{1}{k} \, q \, \big( h; A, \a, \b \big)
\ee
Therefore if a local solution of (\ref{quintic}) exists for a given $(\alpha,\beta)$ within a certain radial interval, it would also be present for $(k \alpha,k^2 \beta)$, for $k>0$, with $h$ being replaced by $h/k$ and the radial interval rescaled by $1/\sqrt[3]{k}$. As a result, each point belonging to the $\a>0$ part of the parabola $\b = c \, \aq$ of the phase space (with $c$ any non-vanishing constant) shares the same physics, hence having the same number of global solutions and matching properties. The same is true for the points belonging the $\a<0$ part of the parabola. So, to understand the global structure of the phase space, it is suf\mbox{}f\mbox{}icient to analyse one point for each of the half parabolas present in the phase space.

It is worthwhile to point out that our starting equations (\ref{solfab})-(\ref{quintic}) were
constructed assuming $GM < \rho < 1/m$, but in the following analysis we use the whole radial
domain $0 < \r < + \infty$ . On one hand, this allows us to characterise exactly the number and
properties of solutions on large and small scales. On the other hand, the picture we have in
mind is that the Compton wavelength of the gravitational f\mbox{}ield $\r_c = 1/m$ is of the same order of the Hubble radius today, and that there is a huge hierarchy between $\r_c$ and the gravitational radius\footnote{We are using units where the speed of light speed has unitary value.} $\r_g = GM$, \textit{i.e.} $\r_c / \r_g \ggg 1$. Therefore, we expect that extending the analysis to the whole radial domain captures the correct physical results.

\subsection{Asymptotic and inner solutions}

We sum up here the results obtained in the appendix on the existence and properties of asymptotic and inner
solutions of eq.~(\ref{quintic}). We refer the reader to the appendix for the details.

\subsubsection{Asymptotic solutions}

In a neighbourhood of $\r \to +\infty$ there are, depending on the value of $(\a, \b)$, three or f\mbox{}ive solutions
to eq.~(\ref{quintic}). In particular:
\begin{itemize}
 \item[-] There is always a decaying solution, which we indicate with $\textbf{L}$. Its asymptotic behaviour is
          \be
          h(\r) = - \frac{2}{3} \left( \frac{\rho_{v}}{\r} \right)^{\! 3} + \, R(\r)
          \ee
          where $\lim_{\r \to +\infty} \, \r^3 R(\r) = 0$. This solution corresponds to a spacetime which is
          asymptotically f\mbox{}lat, as one can see from eqs.~(\ref{solfab})-(\ref{solnab}).
 \item[-] Additionally, there are two or four solutions to eq.~(\ref{quintic}) which tend to a f\mbox{}inite,
          nonzero value as $\r \to +\infty$. We name these solutions with $\textbf{C}_{+}$, $\textbf{C}_{-}$,
          $\textbf{P}_{1}$ and $\textbf{P}_{2}$ (details about this denomination are given in the appendix). Their asymptotic behaviour is
          \be
          h(\r) = C + \, R(\r)
          \ee
          where $\lim_{\r \to +\infty} \, R(\r) = 0$ and $C$ is a root of the reduced asymptotic equation
	  (\ref{red asymptotic equation}). From eqs.~(\ref{solfab})-(\ref{solnab}), one can get convinced that these solutions correspond to spacetimes which are asymptotically non-f\mbox{}lat. Interestingly, the leading term in the gravitational potentials scales as $\rho^2$ for large radii, the same scaling which we f\mbox{}ind in a de Sitter spacetime. It is worthwhile to point out that, since we are working on scales below the Compton wavelength of the gravitational f\mbox{}ield, ``asymptotically non-f\mbox{}lat'' really means that (from the non-truncated theory point of view) the spacetime correspondent to this solution tends to a non-f\mbox{}lat spacetime when the Compton wavelength is approached. To understand the ``true'' asymptotic behaviour of this solution, one should use the non-truncated equations. Note that, even if $C$ (and so $h$) is much smaller than one, the gravitational potentials $n$ and $f$ can be very large (as they behave like $\propto \rho^2$ far from the origin in this case): therefore, the linear approximation of the non-truncated theory we used to obtain eqs.~(\ref{linsoln})-(\ref{linsolf}) is not valid. Instead, the asymptotic fate of the solution is dictated by the nonlinear behaviour of the non-truncated equations. This seems not too easy to predict without a separate analysis, and we don't attempt to address this interesting problem in the present paper.
\end{itemize}

\subsubsection{Inner solutions}

In a neighbourhood of $\r \to 0^+$ there are either one or three solutions to eq.~(\ref{quintic}).
For $\b > 0$ there are exactly three inner solutions, while for $\b < 0$ there is only one inner solution.
In particular:
\begin{itemize}
 \item[-] There is always a diverging solution, which we denote by $\textbf{D}$. Its leading behaviour is
          \be
          h(\r) = - \, \sqrt[3]{\frac{2}{\b}} \, \frac{\r_v}{\r} + R(\r)
          \ee
          where $\lim_{\r \rightarrow 0^+} \, (R(\r)/\r)$ is f\mbox{}inite. This solution exists for both
          $\b > 0$ and $\b < 0$, with opposite signs for each case.
          Using this solution in eqs.~(\ref{solfab})-(\ref{solnab}), one realises that the $h^3$ term cancels the $GM/\rho$ term, so the gravitational f\mbox{}ield is self-shielded and does not diverge as $\r \to 0^+$. This solution is in strong disagreement with gravitational observations.

 \item[-] For $\b > 0$, there are two additional solutions to eq.~(\ref{quintic}), which tend to a f\mbox{}inite, non-zero value as $\r \to 0^+$.
          We indicate these solutions by $\textbf{F}_{+}$ and $\textbf{F}_{-} \,$. Their leading behaviour is
          \be
          h(\r) = \pm \sqrt{\frac{1}{3 \, \b}} + \, R(\r)
          \ee
          where $\lim_{\r \to 0^+} \, R = 0$. Notice that for $\b < 0$ there are no solutions to eq.~(\ref{quintic}) which tend to a f\mbox{}inite value as $\r \to 0^+$.

          The expressions (\ref{solfab})-(\ref{solnab}) for the gravitational potentials imply that the metric associated to these solutions ($\textbf{F}_{+}$ and $\textbf{F}_{-}$) approximate the linearised Schwarzschild metric as $\r \to 0^+$.
\end{itemize}
From the behaviour of the inner solutions, one concludes that only in the $\b>0$ part of the phase space solutions may exhibit the Vainshtein mechanism, but not necessarily for all values of $\alpha$. In the next subsection we see more in detail how this mechanism works.

\subsubsection{Vainshtein mechanism}

In order to study where in the phase space the Vainshtein mechanism works, it is useful to compare the
gravitational potentials $f$ and $n$ with their counterparts in the GR case. In the weak f\mbox{}ield limit, the Schwarzschild solution of GR reads
\be
ds^2 = - \bigg( 1 - \frac{2 G M}{\r} \bigg) \, dt^2 + \bigg( 1 + \frac{2 G M}{\r} \bigg) \, d\rho^2
+ \rho^2 \, d \Omega^2
\ee	
so by calling $f_{GR} = n_{GR} = - 2 G M / \r$ we obtain
\begin{align}
\label{solfabderx}
\frac{f}{f_{GR}} &= 1 + \frac{1}{2} \, \bigg( \frac{\rho}{\rho_{v}} \bigg)^3 \, \Big( h - \alpha h^2 + \beta h^3 \Big) \\[2mm]
\label{solnabderx}
\frac{n^{\, \p}}{n_{GR}^{\, \p}} &= 1 - \frac{1}{2} \, \bigg( \frac{\rho}{\rho_{v}} \bigg)^3 \, \Big( h - \b h^3 \Big)
\end{align}
Let us now f\mbox{}irst discuss the asymptotic solutions. For the decaying solution $\textbf{L}$, we have that the linear
contribution in $h$ rescales the coef\mbox{}f\mbox{}icients of the Schwarzschild-like terms, so we obtain
$f / f_{GR} \to 2/3$ and $n^{\, \p} / n_{GR}^{\, \p} \to 4/3$ for $\r \to +\infty$.
For the non-decaying solutions $\textbf{C}_{\pm}$ and  $\textbf{P}_{1,2}$,
the leading behaviour for $f / f_{GR}$ and $n^{\, \p} / n_{GR}^{\, \p}$ is proportional
to $( \r / \r_v )^3$ in both cases, however the proportionality coef\mbox{}f\mbox{}icients generally
dif\mbox{}fer since they have a dif\mbox{}ferent functional dependence on $\alpha$ and $\beta$. There are some
special cases for $(\a,\b)$ where these asymptotic solutions lead to $f/n\to 1$ as $\rho\to+\infty$, and therefore have the same behaviour as in a de Sitter spacetime.

Consider instead the inner solutions. For the f\mbox{}inite solutions $\textbf{F}_{\pm}$ we obtain
$(f / f_{GR}) \to 1$ and $(n^{\, \p} / n_{GR}^{\, \p}) \to 1$ as $\r \to 0^+$,
where the corrections scale like $\r^3$. On the contrary, for the diverging solution $\textbf{D}$,
the cubic terms in $h$ cancel out the contribution coming from the Schwarzschild-like terms, as
explained above, and so $(f / f_{GR}) \to 0$ and $(n^{\, \p} / n_{GR}^{\, \p}) \to 0$ when
$\r \to 0^+$. In this case, corrections are linear in $\r$.

Therefore, any global solution of equation (\ref{quintic}) which interpolates between $\textbf{L}$ and
$\textbf{F}_{\pm}$ provides a realisation of the Vainshtein mechanism in an asymptotically f\mbox{}lat
spacetime, whereas an interpolation between $\textbf{C}_{\pm}$ or $\textbf{P}_{1,2}$ with
$\textbf{F}_{\pm}$ exhibits the Vainshtein mechanism in an asymptotically non-f\mbox{}lat spacetime.
Furthermore, notice that any asymptotic solution which interpolates with the inner solution
$\textbf{D}$ does no lead to the Vainshtein mechanism. These matchings will be explicitly
exposed in the next section.

\section{Results}

\subsection{Solutions matching}

\begin{figure}[htp!]
\begin{center}
\includegraphics[width=12cm]{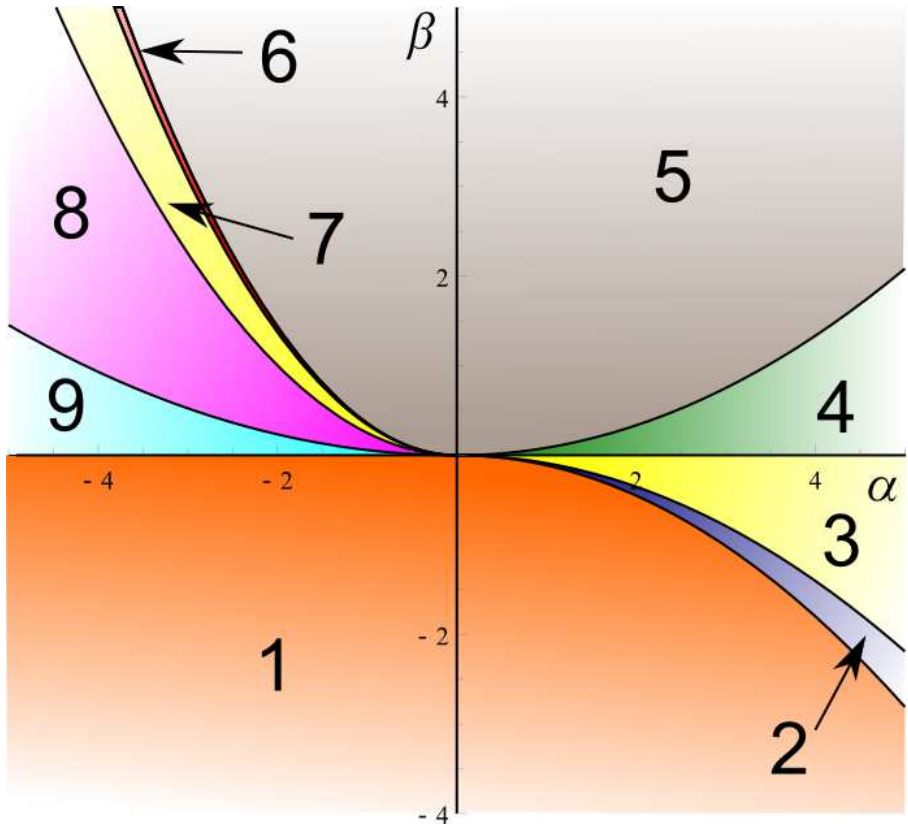}
\caption{Phase space diagram in $(\a,\b)$ for the solutions to the quintic equation
(\ref{quintic}) in $h$, where the dif\mbox{}ferent regions show dif\mbox{}ferent matching of inner
solutions to asymptotic ones. The lines splitting the regions are half parabolas
($\b \propto \aq$, with $\a >0$ or $\a <0$) due to rescaling symmetry of eq.~(\ref{quintic}).}
\label{phase space}
\end{center}
\end{figure}

The phase space diagram which displays our results about solution matching is given in f\mbox{}igure
\ref{phase space}. We discuss separately the $\b > 0$ and $\b < 0$ part of the phase space, and
refer to the f\mbox{}igure for the numbering of the regions. The notation $\textbf{I}
\leftrightarrow \textbf{A}$ means that there is matching between the inner solution $\textbf{I}$
and the asymptotic solution $\textbf{A}$.
\\
\\
\small
\textbf{$\b < 0$}
\normalsize
\\
\\
In this part of the phase space, there is only one inner solution, $\textbf{D}$, so there can be at most one global solution to (\ref{quintic}). There are three distinct regions which dif\mbox{}fer in the way the matching works:
\begin{itemize}
 \item[-] region 1: $\textbf{D} \leftrightarrow \textbf{C}_{+}$. In this region, there are three or f\mbox{}ive asymptotic solutions, and only one of them, $\textbf{C}_{+}$, is positive. This solution is the one which connects with the inner solution $\textbf{D}$, which is also positive, leading to the only global solution of eq.~(\ref{quintic}). The boundaries of this region are the line $\b = 0$ for $\a<0$ and the parabola $\b = c_{12} \, \a^2$ for $\a>0$, where $c_{12}$ is the negative\footnote{The equation $-4 - 8 \, y + 88 \, y^2 - 1076 \, y^3 + 2883 \, y^4 = 0$ has only two real roots, one positive and one negative.} root of the equation $-4 - 8 \, y + 88 \, y^2 - 1076 \, y^3 + 2883 \, y^4 = 0$ (approximatively, $c_{12} \simeq -0.1124$). On the boundary $\b = c_{12} \, \a^2$ the matching $\textbf{D} \leftrightarrow \textbf{C}_{+}$ still holds, however the solution $h(\r)$ displays an inf\mbox{}lection point with vertical tangent.

 \item[-] region 2: $\textbf{No matching}$. In this region there are three asymptotic solutions. However, none of them can be extended all the way to $\r \to 0^+$, and so, despite the fact that local solutions exist both at inf\mbox{}inity and near the origin, equation (\ref{quintic}) does not admit any global solution. The boundaries of this region are the parabola $\b = c_{12} \, \a^2$ and the (negative) f\mbox{}ive-roots-at-inf\mbox{}inity parabola $\b = c_{-} \, \a^2$, where $c_{-}$ is the only real root of the equation $8 + 48 \, y - 435 \, y^2 + 676 \, y^3 = 0$ (approximatively, $c_{-} \simeq -0.0876$).

 \item[-] region 3: $\textbf{D} \leftrightarrow \textbf{P}_{2}$. This region coincides with the $\a > 0$,
$\b < 0$ part of the f\mbox{}ive roots at inf\mbox{}inity region of the phase space (see f\mbox{}ig.
\ref{five roots}). The largest positive asymptotic solution, $\textbf{P}_{2}$, is the one which connects to $\textbf{D}$, leading to the only global solution of eq.~(\ref{quintic}). On the boundary $\b = c_{-} \, \a^2$ the matching $\textbf{D} \leftrightarrow \textbf{P}_{2}$ still holds, but the solution $h$ seen as a function of $A$ has inf\mbox{}inite derivative in $A=0$.
 \\
 \end{itemize}

\noindent \small
          \textbf{$\b > 0$}
          \normalsize
\\
\\
In this part of the phase space, there are three inner solutions, $\textbf{D}$, $\textbf{F}_{+}$ and $\textbf{F}_{-}$, so there can be at most three global solutions to eq.~(\ref{quintic}). There are six distinct regions with dif\mbox{}ferent matching properties:
\begin{itemize}

 \item[-] region 4: $\textbf{F}_{-} \leftrightarrow \textbf{L}$ , $\textbf{D} \leftrightarrow \textbf{C}_{-}$. This region lies inside the $\a > 0$, $\b > 0$ part of the f\mbox{}ive roots at inf\mbox{}inity region of the phase space (see f\mbox{}ig.~\ref{five roots}), so there are f\mbox{}ive asymptotic solutions. Of the f\mbox{}ive asymptotic solution, $\textbf{C}_{-}$ and $\textbf{L}$ can always be extended to $\r \to 0^+$, while $\textbf{C}_{+}$, $\textbf{P}_{1}$ and  $\textbf{P}_{2}$ cannot. So there are just two global solutions to eq.~(\ref{quintic}). The boundaries of this region are the parabola $\b = c_{45} \, \a^2$, where $c_{45} = 1/12 \simeq 0.0833$, and the line $\b = 0$. On the boundary $\b = c_{45} \, \a^2$ there is the additional matching $\textbf{F}_{+} \leftrightarrow \textbf{C}_{+}$, and the correspondent solution is $h(\r) = const = + \sqrt{1/\,3\,\b}\,$.

 \item[-] region 5: $\textbf{F}_{+} \leftrightarrow \textbf{C}_{+}$ , $\textbf{F}_{-} \leftrightarrow \textbf{L}$, $\textbf{D} \leftrightarrow \textbf{C}_{-}$. In this region there are three or f\mbox{}ive asymptotic solutions; $\textbf{C}_{-}$ , $\textbf{C}_{+}$ and $\textbf{L}$ can always be extended to $\r \to 0^+$, while $\textbf{P}_{1}$ and $\textbf{P}_{2}$ , where present, cannot. So there are three global solutions to (\ref{quintic}). The boundaries of this region are the parabola $\b = c_{45} \, \a^2$ for $\a > 0$ and the parabola $\b = c_{56} \, \a^2$ for $\a < 0$, where $c_{56} = (5 + \sqrt{13})/24 \simeq 0.3586$. On the $\a < 0$ boundary $\b = c_{56} \, \a^2$ the matching works as in the rest of the region, but the solution $\textbf{F}_{-} \leftrightarrow \textbf{L}$ has an inf\mbox{}lection point with vertical tangent.

 \item[-] region 6: $\textbf{D} \leftrightarrow \textbf{C}_{-}$ , $\textbf{F}_{+} \leftrightarrow \textbf{C}_{+}$. In this region there are three asymptotic solutions, however only two of them can be extended to $\r \to 0^+$, while $ \textbf{L}$ cannot. Therefore, there are just two global solutions to eq.~(\ref{quintic}). The boundaries of this region are the parabolas $\b = c_{56} \, \a^2$ and $\b = c_{67} \, \a^2$, where $c_{67}$ is the positive root of the equation $-4 - 8 \, y + 88 \, y^2 - 1076 \, y^3 + 2883 \, y^4 = 0$ (approximatively, $c_{67} \simeq 0.3423$). On the boundary $\b = c_{67} \, \a^2$ the matching works as in the rest of the region, but the solution $\textbf{D} \leftrightarrow \textbf{C}_{-}$ has an inf\mbox{}lection point with vertical tangent.

 \item[-] region 7: $\textbf{F}_{+} \leftrightarrow \textbf{C}_{+}$. In this region there are three asymptotic solutions, however only one of them can be extended to $\r \to 0^+$, while $\textbf{L}$ and $\textbf{C}_{-}$ cannot. The boundaries of this region are the parabola $\b = c_{67} \, \a^2$ and the (positive) f\mbox{}ive-roots-at-inf\mbox{}inity parabola $\b = c_{+} \, \a^2$, where $c_{+} = 1/4$. Note that on the ($\a < 0$) part of the parabola $\b = 1/3 \, \aq$ there is the additional matching $\textbf{F}_{-} \leftrightarrow \textbf{C}_{-}$, so for these points there are two global solutions to eq.~(\ref{quintic}). On the boundary $\b = c_{+} \, \a^2$ there are the additional matchings $\textbf{F}_{-} \leftrightarrow \textbf{P}_{1}$ , $\textbf{D} \leftrightarrow \textbf{P}_{2}$, and the solutions corresponding to both of these additional matchings, seen as functions of $A$, display an inf\mbox{}inite derivative in $A=0$.

 \item[-] region 8: $\textbf{F}_{+} \leftrightarrow \textbf{C}_{+}$ , $\textbf{F}_{-} \leftrightarrow \textbf{P}_{1}$ , $\textbf{D} \leftrightarrow \textbf{P}_{2}$. This region lies inside the $\a < 0$, $\b > 0$ part of the f\mbox{}ive roots at inf\mbox{}inity region of the phase space (see f\mbox{}ig.~\ref{five roots}), so there are f\mbox{}ive asymptotic solutions. Only three of them can be extended to $\r \to 0^+$, while $ \textbf{C}_{-}$ and $\textbf{L}$ cannot. The boundaries of this region are the parabolas $\b = c_{+} \, \a^2$ and $\b = c_{89} \, \a^2$, where $c_{89} = (5 - \sqrt{13})/24 \simeq 0.0581$. On the boundary $\b = c_{89} \, \a^2$ the matchings are the same as in the rest of the region, but the solution $h(\r)$ correspondent to the matching $\textbf{F}_{+} \leftrightarrow \textbf{C}_{+}$ has an inf\mbox{}lection point with vertical tangent.

 \item[-] region 9: $\textbf{F}_{-} \leftrightarrow \textbf{P}_{1}$ , $\textbf{D} \leftrightarrow \textbf{P}_{2}$. This region lies inside the $\a < 0$, $\b > 0$ part of the f\mbox{}ive roots at inf\mbox{}inity region of the phase space (see f\mbox{}ig.~\ref{five roots}), so there are again f\mbox{}ive asymptotic solutions. The matching is similar to that of region 8, apart from the fact that $\textbf{C}_{+}$ cannot be extended to $\r \to 0^+$ anymore; hence there are just two global solutions to eq.~(\ref{quintic}). The boundaries of this region are the parabola $\b = c_{89} \, \a^2$ and line $\b = 0$.
\end{itemize}

\noindent We note that the decaying solution $\textbf{L}$ never connects to the diverging one $\textbf{D}$,
so we cannot have a spacetime which is asymptotically f\mbox{}lat and exhibit the self-shielding of the
gravitational f\mbox{}ield at the origin. On the other hand, f\mbox{}inite non-zero asymptotic solutions ($\textbf{C}_{\pm}$ or $\textbf{P}_{1,2}$) can connect to both f\mbox{}inite and diverging inner solutions.
Therefore, one can have an asymptotically non-f\mbox{}lat spacetime which presents self-shielding at the origin, or an asymptotically non-f\mbox{}lat spacetime which tends to Schwarzschild spacetime for small radii. More precisely, for $\b < 0$ there are only solutions displaying the self-shielding of the gravitational f\mbox{}ield, apart from region 2 where there are no global solutions. Therefore the Vainshtein mechanism never works for $\b < 0$. In contrast, for $\b > 0$ all three kinds of global solutions are present. Solutions with asymptotic f\mbox{}latness and the Vainshtein mechanism are present in regions 4 and 5, while solutions which are asymptotically non-f\mbox{}lat and exhibit the Vainshtein mechanism do exist in all ($\b > 0$) regions but region 4. Finally, solutions which display the self-shielding of the gravitational f\mbox{}ield are present in all ($\b > 0$) regions but region 7.

\subsection{Numerical solutions}

\begin{figure}[htp!]
\begin{center}
\includegraphics[width=11cm]{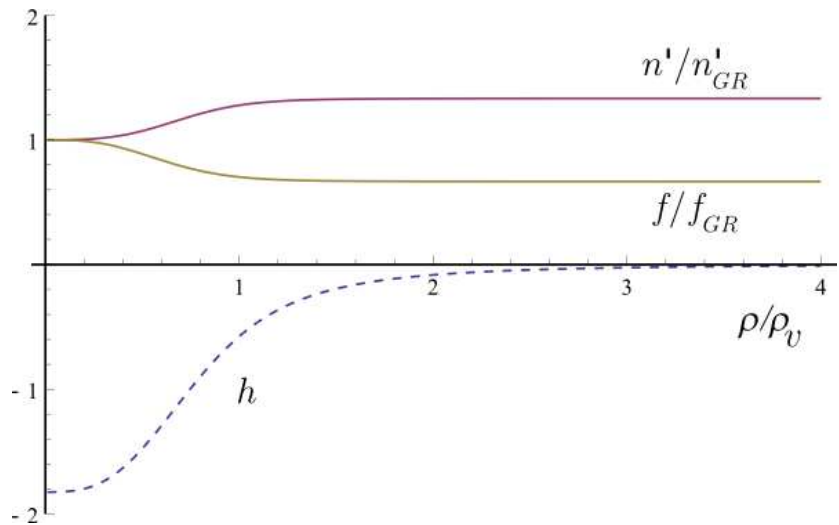}
\caption{Numerical solutions for the case $\textbf{F}_{-} \leftrightarrow \textbf{L}$.}
\label{Numerical1}
\end{center}
\end{figure}

We present here the numerical solutions for the $h$ f\mbox{}ield and the gravitational potentials in some
representative cases. We choose a specif\mbox{}ic realisation for each of the three physically distinct
cases, namely asymptotic f\mbox{}latness with Vainshtein mechanism, asymptotically non-f\mbox{}lat spacetime with Vainshtein mechanism, and asymptotically non-f\mbox{}lat spacetime with self-shielded gravitational f\mbox{}ield at the origin. In addition, we consider the case in which there are no global solutions to eq.~(\ref{quintic}). This provides an illustration of what happens, in general, to local solutions of eq.~(\ref{quintic}) which cannot be extended to the whole radial domain, and give an insight on the phenomenology of equation (\ref{quintic}).
\\
\\
\small
\textbf{Asymptotic f\mbox{}latness with Vainshtein mechanism}
\normalsize
\\
\\
Let's consider the case in which the solution of eq.~(\ref{quintic}) connects to the decaying solution
at inf\mbox{}inity $\textbf{L}$ and to a f\mbox{}inite inner solution (in this case
$\textbf{F}_{-}$). In f\mbox{}igure \ref{Numerical1}, the
numerical solutions for $h$ (dashed line), $f / f_{GR}$ (bottom continuous line) and
$n^{\, \p} / n_{GR}^{\, \p}$ (top continuous line) are plotted as functions of
the dimensionless radial coordinate $x \equiv \r / \r_v$. These solutions correspond to the point
$(\alpha, \beta) = (0 \, , 0.1)$ of the phase space.

\begin{figure}[htp!]
\begin{center}
\includegraphics[width=11cm]{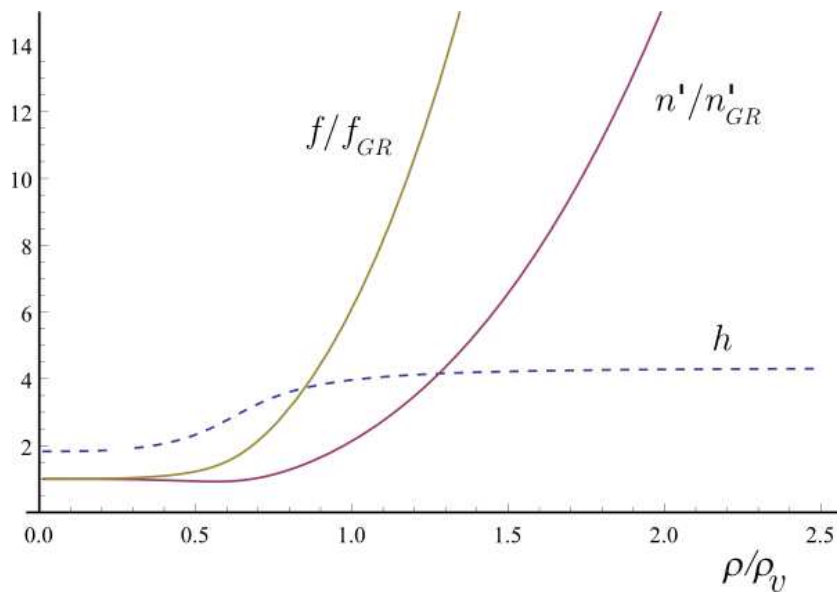}
\caption{Numerical solutions for the case $\textbf{F}_{+} \leftrightarrow \textbf{C}_{+}$.}
\label{Numerical2}
\end{center}
\end{figure}

This plot displays very clearly the presence of the vDVZ discontinuity and its resolution \emph{via}
the Vainshtein mechanism. For large scales, $h$ is small and the gravitational potentials behave like
the Schwarzschild one, however their ratio is dif\mbox{}ferent from one, unlike the massless case.
Note that the ratio of the two potentials for $\r \gg \r_v$ is independent of $m$, so does not
approach one as $m \to 0$ (vDVZ discontinuity). However, on small scales $h$ is strongly coupled, and
well inside the Vainshtein radius the two potentials scale again as the Schwarzschild one, but their
ratio is now one even if $m \neq 0$. So, the strong coupling of the $h$ f\mbox{}ield on small scales restores
the agreement with GR (Vainshtein mechanism).
\\
\\
\small
\textbf{Asymptotically non-f\mbox{}lat spacetime with Vainshtein mechanism}
\normalsize
\\
\\
Let's consider now the case in which the solution of eq.~(\ref{quintic}) connects to a f\mbox{}inite solution at
inf\mbox{}inity and to a f\mbox{}inite inner solution. We consider for def\mbox{}initeness the phase space
point $(\alpha, \beta) = (0 \, , 0.1)$. In f\mbox{}igure \ref{Numerical2}, we plot the numerical results
for the gravitational potentials
(normalised to their GR values) and the global solution of eq.~(\ref{quintic}) which interpolates between
the inner solution $\textbf{F}_{+}$ and the asymptotic solution $\textbf{C}_{+}$.

\begin{figure}[htp!]
\begin{center}
\includegraphics[width=11cm]{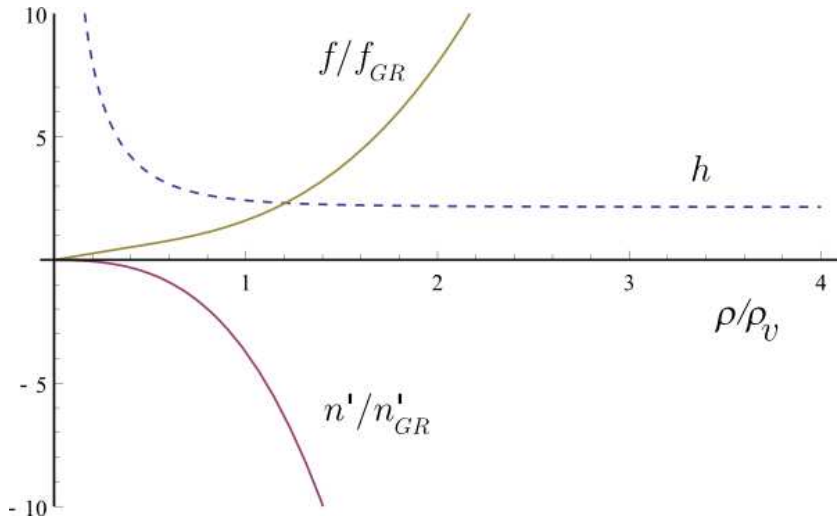}
\caption{Numerical solutions for the case $\textbf{D} \leftrightarrow \textbf{C}_{+}$.}
\label{numerical3}
\end{center}
\end{figure}

We can see that, on large scales, the gravitational potentials are not only dif\mbox{}ferent one from the
other but also behave very dif\mbox{}ferently compared to the GR case. However, on small scales there is a macroscopic region where the two potentials agree, and their ratio with the Schwarzschild potential stays nearly constant and equal to one. Therefore, also in this case the small scale behaviour of $h$ guarantees that GR results are recovered, even if the spacetime is not asymptotically f\mbox{}lat. This behaviour provide then, in a more general sense, a realisation of the Vainshtein mechanism.
\\
\\
\small
\textbf{Asymptotically non-f\mbox{}lat spacetime with self-shielding}
\normalsize
\\
\\
We turn now to the case where the solution of eq.~(\ref{quintic}) connects to a f\mbox{}inite solution at inf\mbox{}inity and to the diverging inner solution. In f\mbox{}igure \ref{numerical3}, we plot the global solution $h$ and the associated gravitational potentials, normalised to their GR values, correspondent to the phase space point $(\alpha, \beta) = (-1 \, , -0.5)$. It is apparent that there are no regions where the solutions behave like in the GR case.

\begin{figure}[htp!]
\begin{center}
\includegraphics[width=11cm]{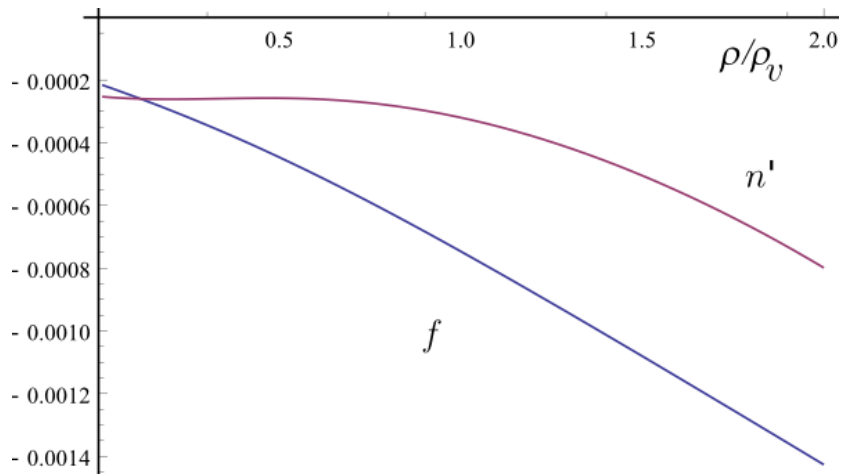}
\caption{Numerical solutions for the gravitational potentials, for the case
$\textbf{D} \leftrightarrow \textbf{C}_{+}$.}
\label{numerical3nonnorm}
\end{center}
\end{figure}

\begin{figure}[htp!]
\begin{center}
\includegraphics[width=11cm]{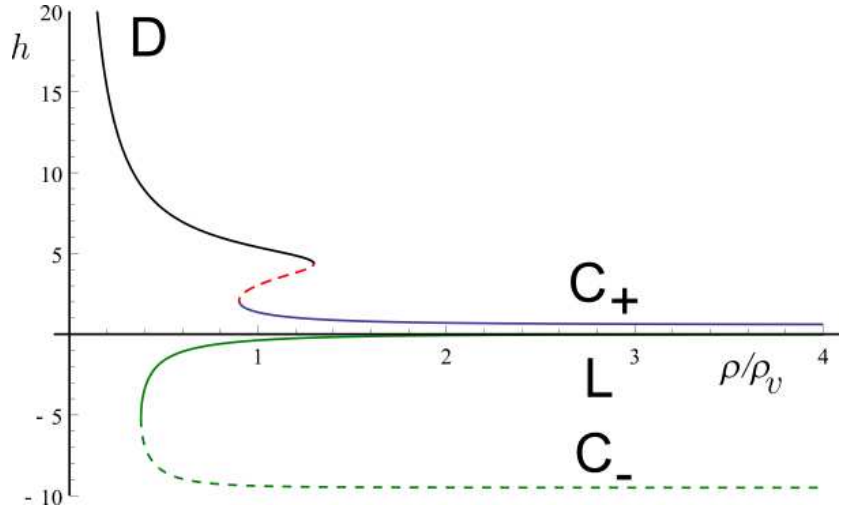}
\caption{Numerical results for all local solutions of eq.~(\ref{quintic}) in the case where there is no
matching.}
\label{Numerical4}
\end{center}
\end{figure}

To see that the gravitational potentials are indeed f\mbox{}inite at the origin, we plot in
f\mbox{}igure \ref{numerical3nonnorm} the potentials $f$ and $n'$ themselves, as functions
of $\rho/\r_v$. We choose for def\mbox{}initeness the following ratio between the
Compton wavelength and the gravitational radius $\r_{c} / \r_{g} = 10^{6}$, and plot
the potentials for $0.01 < \rho/\r_v < 2$. Note that, since in this case $\r_{c} / \r_{v}
= \sqrt[3]{\r_{c} / \r_{g}} = 10^{2}$, the range where the functions are plotted
is well inside the range of validity of our approximations. We can see that the
potentials approach a f\mbox{}inite value as $\r \to 0^+$, and so indeed the gravitational
f\mbox{}ield does not diverge at the origin.
\\
\\
\small
\textbf{No matching}
\normalsize
\\
\\
Finally, we consider the case in which equations (\ref{solfab}) $-$ (\ref{quintic}) do not admit global
solutions. We consider for def\mbox{}initeness the phase space point $(\alpha, \beta) = (1 \, , -0.092)$.
In f\mbox{}igure \ref{Numerical4} we plot all the local solutions of the quintic equation (\ref{quintic})
as functions of the dimensionless radial coordinate $x \equiv \r / \r_v$.

For $0 < x < 0.4$, there is only one local solution (the top continuous curve), which connects to the diverging inner solution $\textbf{D}$. At $x \simeq 0.4$ a pair of solutions is created (dashed and continuous negative valued curves), and at $x \simeq 0.9$, another pair of solutions is created (positive valued dashed curve and positive valued bottom continuous curve). However, at $x \simeq 1.3$ one of the newly created functions (the positive valued dashed curve) annihilates with the solution with connects to the inner solution, so for $x > 1.3$ there are three local solutions, which f\mbox{}inally connect with the asymptotic solutions $\textbf{C}_{-}$ , $\textbf{L}$ and $\textbf{C}_{+}$. Therefore, the number of existing local solutions is one for $0 < x < 0.4$, three for $0.4 < x < 0.9$, f\mbox{}ive for $0.9 < x < 1.3$ and three for $x > 1.3$. We can see that, despite the fact that for every $\r$ there is at least one local solution, there does not exist a solution which extends over the whole radial
domain. Note that the solutions are created and annihilated in pairs; furthermore, the pairs of solutions have inf\mbox{}inite slope when they are created and when they annihilate, which results in the gravitational potentials having diverging derivatives when the creation/annihilation points are approached. These are found to be general features of the phenomenology of equation (\ref{quintic}). In fact, in most part of the phase space some of the asymptotic/inner solutions cannot be extended to the whole radial domain $0 < \r < +\infty$. However, it never happens that the value of the local solutions diverges at a f\mbox{}inite radius, while they always disappear or are created in pairs, with their values remaining bounded but with derivatives which diverge.

\section{Conclusions}

Recently, a ghost-free  model of  non-linear massive gravity has been  proposed. We studied static, spherically symmetric solutions of this theory inside the Compton radius of the gravitational f\mbox{}ield, and considered the weak f\mbox{}ield limit for the gravitational potentials, while keeping all non-linearities of the helicity-0 mode. For every point of the two free parameter phase space, we characterised completely the number and properties of asymptotic solutions on large scales and also of inner solutions on small scales. In particular, there are two kinds of asymptotic solutions, where one of them is asymptotically f\mbox{}lat and the other one is not. There are also two kinds of inner solutions, one which displays the Vainshtein mechanism and the other which exhibits the self-shielding of the gravitational f\mbox{}ield near the origin.

We described under which circumstances the theory admits global solutions interpolating between the asymptotic and inner solutions, and found that the asymptotically f\mbox{}lat solution connects only to inner solutions displaying the Vainshtein mechanism, while solutions which diverge asymptotically can connect to both kinds of inner solutions.
Furthermore, we showed that there are some regions in the parameter space where global solutions do not exist, and characterised precisely in which regions of the phase space the Vainshtein mechanism is working.

Our study embraces all of the phase space spanned by the two parameters of the theory. Notably, we found that, within our approximations, the asymptotic and inner solutions cannot in general be extended to the whole radial domain. In particular, we exhibited  extreme cases in which  global solutions do not exist at all. This happens because at a f\mbox{}inite radius the derivatives of the metric components diverge, while the metric components themselves remain bounded. When the derivatives of the metric cease to be small, the approximations we used to derive the equations (\ref{solfab}) $-$ (\ref{quintic}) break down. It would be interesting to study what happens at this radius in the full theory.

\section*{Acknowledgments}
GN and KK are supported by the European Research Council. KK is also supported by the STFC (grant no. ST/H002774/1) and the Leverhulme trust. GT is supported by an STFC Advanced Fellowship ST/H005498/1.

\newpage

\begin{appendix}

\section{Asymptotic behaviour of solutions of the quintic equation}

In this appendix we study the local solutions of the quintic equation (\ref{quintic}) in a
neighbourhood of $\rho = 0^+$ and also in a neighbourhood of $\rho = +\infty$.
As previously mentioned we consider only the $\b \neq 0$ case, since for $\b = 0$ the equation
becomes a cubic and it is possible to f\mbox{}ind all local solutions analytically (see \cite{us2}).

\subsection{Asymptotic and inner solutions}

\subsubsection*{Asymptotic solutions}

Suppose that a solution $h(\r)$ of the quintic equation (\ref{quintic}), which for convenience
we rewrite here,
\be
\label{quinticAppendix}
\frac{3}{2} \, \bq \, \ha{5} - \Big( \aq + 2 \b \Big) \, \ha{3} + 3 \, \Big( \a + \b A \Big) \, \ha{2}
- \frac{3}{2} \, h - A = 0
\ee
exists in a neighbourhood of $\rho = + \infty \,$, and that it has a well def\mbox{}ined limit as
$\r \rightarrow + \infty$. Then such a solution cannot be divergent. To see this, suppose that
indeed the solution is divergent $|\lim_{\r \rightarrow + \infty} h(\r)| = + \infty \,$: it is
then possible to divide the quintic equation by $\ha{5}$ (in a neighbourhood of $\rho = + \infty \,$),
obtaining the following equation in $v = 1/h$
\be
\label{vequation}
\frac{3}{2} \, \bq - \big( \aq + 2 \b \big) \, \va{2} + 3 \, \big( \a + \b A \big) \, \va{3} - \frac{3}{2}
\, \va{4} - A \, \va{5} = 0
\ee
Taking the $\rho \to + \infty$ limit of both sides of this expression one obtains $\b = 0$,
which is precisely against our initial assumption.

\begin{figure}[htp!]
\begin{center}
\includegraphics{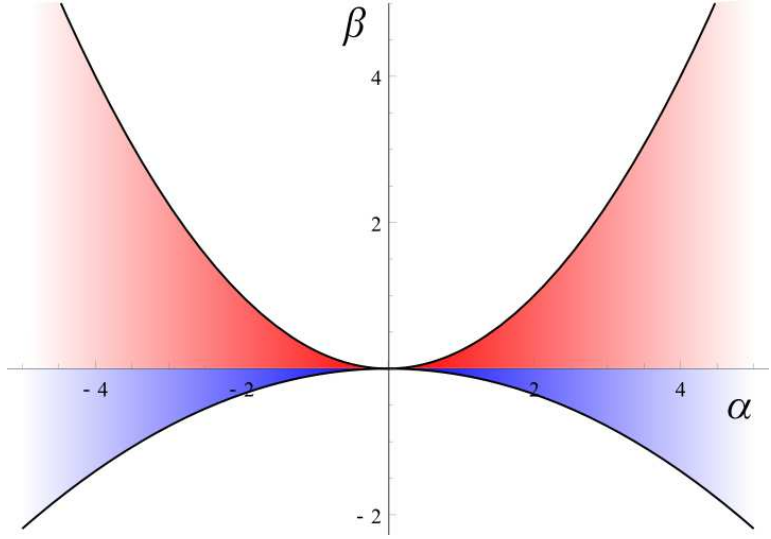}
\caption{phase space diagram for the number of asymptotic solutions}
\label{five roots}
\end{center}
\end{figure}

Suppose now that $\lim_{\r \rightarrow + \infty} h(\r)$ is f\mbox{}inite, and let's call it $C$. Then
both of the sides of the quintic equation itself have a f\mbox{}inite limit when
$\rho \to + \infty \,$, and taking this limit one gets
\be
\frac{3}{2} \, \bq \, C^{5} - \big( \aq + 2 \b \big) \, C^{3} + 3 \, \a \, C^{2} - \frac{3}{2} \, C = 0
\ee
It follows then that the allowed asymptotic values at inf\mbox{}inity for $h(\r)$ are the roots of the
following equation, which we call the \emph{asymptotic equation}
\be
\label{asymptotic equation}
\mathscr{A}(y) \equiv \frac{3}{2} \, \bq \, y^{5} - \big( \aq + 2 \b \big) \, y^{3} + 3 \, \a \, y^{2}
- \frac{3}{2} \, y = 0
\ee
Note that $y = 0$ is always a root of this equation, and in fact a simple root (\textit{i.e.} a root
of multiplicity one) since $\frac{d}{dy} \mathscr{A}(0) \neq 0 \,$. Dividing by $y$, one obtains
that the other asymptotic values for $h(\r)$ are the roots of the \emph{reduced asymptotic equation}
\be
\label{red asymptotic equation}
\mathscr{A}_{r}(y) \equiv \frac{3}{2} \, \bq \, y^{4} - \big( \aq + 2 \b \big) \, y^{2}
+ 3 \, \a \, y - \frac{3}{2} = 0
\ee
This last equation is a quartic, so it can have up to 4 (real) roots, depending on the specif\mbox{}ic
values of $\a$ and $\b$. Since $\lim_{h \rightarrow \pm \infty} \mathscr{A}_{r}(h) =
\pm \infty$ and $\mathscr{A}_{r}(0) = - 3/2 < 0 \,$, it has always at least two roots, one positive
and one negative. For the same reason, it cannot have two positive and two negative roots, since
at each simple root the quartic function changes sign. Since the function
$\mathscr{A}(h)$ changes smoothly with $\a$ and $\b$, the boundaries between regions where there
are f\mbox{}ive roots and regions where there are three roots are found enforcing that $\mathscr{A}(h)$
has a multiple root. This condition is satisf\mbox{}ied only by the points belonging to
the parabolas $\b = c_{+} \, \a^2$ and $\b = c_{-} \, \a^2$, where $c_{+} = 1/4$ and $c_{-}$ is the only
real root of the equation $8 + 48 \, y - 435 \, y^2 + 676 \, y^3 = 0$. The regions above the positive
parabola and below the negative one have only three roots, which are simple roots, while the regions between
the two parabolas (except $\b = 0$) have f\mbox{}ive roots, which are again simple roots. On the boundaries $\b
= c_{\pm} \, \a^2$ between the three-roots regions and the f\mbox{}ive-roots regions there are four roots, one
of which is a root of multiplicity two. This is summarised in f\mbox{}igure \ref{five roots}.

We name the roots in the following way: the $y=0$ root is denoted as $\textbf{L}$. For the phase space
points where there are just three roots, the positive root is denoted as $\textbf{C}_{+}$ and the
negative one as $\textbf{C}_{-}$ . For points in the f\mbox{}ive-roots regions, we adopt the following
convention. Be $(\a_5, \b_5)$ a point where there are f\mbox{}ive roots. In the same quadrant of the
phase space, take another point $(\a_3, \b_3)$ where there are three roots, and a path $\mathscr{C}$ which
connects the two points.
Following the path $\mathscr{C}$, two of the four non-zero roots of $(\a_5, \b_5)$ smoothly f\mbox{}low to the non-zero roots of
$(\a_3, \b_3)$, and are denoted as $\textbf{C}_{+}$ and $\textbf{C}_{-}$ themselves. The other two
non-zero roots of $(\a_5, \b_5)$, instead, disappear when (following $\mathscr{C}$) the boundary of the
f\mbox{}ive-roots region is crossed, and are denoted as $\textbf{P}_{1}$ and $\textbf{P}_{2}$. We adopt the
convention that $\abs{\textbf{P}_{1}} \leq \abs{\textbf{P}_{2}}$. The def\mbox{}inition is independent of the
particular choice of the point $(\a_3, \b_3)$ and of the path $\mathscr{C}$ used. A careful study
of the asymptotic equation and of its derivatives permits to show that we have
$\textbf{C}_{-} < \textbf{C}_{+} < \textbf{P}_{1} < \textbf{P}_{2}$ for $\a > 0$ and $\textbf{P}_{2} <
\textbf{P}_{1} < \textbf{C}_{-} < \textbf{C}_{+}$ for $\a < 0$. On the boundaries $\b
= c_{\pm} \, \a^2$ we have $\textbf{P}_{1} = \textbf{P}_{2} \equiv \textbf{P}$.

\subsubsection*{Inner solutions}

Suppose now that a solution of the quintic equation exists in a neighborhood of $\r = 0^+$
(possibly not def\mbox{}ined in $\r = 0$), and that it has a well def\mbox{}ined limit when $\r \rightarrow 0^+$.
Then such a solution cannot tend to zero as $\r \to 0^+$. In fact, for $\r \neq 0$ we
can divide (\ref{quinticAppendix}) by $A$, and indicating $x \equiv \r / \r_v$ we obtain
\be
x^3 \, \bigg( \frac{3}{2} \, \bq \, \ha{5} - \big( \aq + 2 \b \big) \, \ha{3} + 3 \, \a \, \ha{2}
- \frac{3}{2} \, h \bigg) + 3 \, \b \, \ha{2} - 1 = 0
\ee
If we had $h \to 0$ as $\r \to 0^+$, then we would get $-1=0$ when taking the $\r \to 0^+$ limit in the previous equation. Therefore, $\lim_{\r \to 0^+} h (\r) \neq 0$ . In a neighbourhood of $\r = 0^+ \,$, we can further divide
the equation above by $\ha{5}$, obtaining the following equation in $v = 1 / h$
\be
\label{quinticv}
\va{5} + \frac{3}{2} \, x^3 \, \va{4} - 3 \, \big( \b + \a \, x^3 \big) \, \va{3} + \big( \aq + 2 \b \big)
\, x^3 \, \va{2} - \frac{3}{2} \, \bq \, x^3 = 0
\ee
The permitted limiting values for $v$ are then the roots of the equation obtained from the previous
one setting $x = 0$, namely
\be
\label{inner equation}
\va{5} - 3 \, \b \, \va{3} = 0
\ee
For $\b > 0$ there are three roots, namely $v_{0} = 0$, $v_{+} = + \sqrt{3 \, \b}$ and
$v_{-} = - \sqrt{3 \, \b}$ ; for $\b < 0$, instead, there is only the root $v = 0$.
Therefore, the permitted limiting behaviours for $h$ when $\r \rightarrow 0^+$ are
\be
\abs{h(\r)} \to + \infty
\ee
for $\b \neq 0$, and
\be
h \to \pm \sqrt{\frac{1}{3 \, \b}}
\ee
only for $\b > 0$.

\subsection{Leading behaviours}

Note that so far we have not proved that inner and asymptotic solutions exist, but just found the
values that have to be the limit of these solutions if they exist. The existence and uniqueness of
solutions can be proved applying the \emph{implicit function theorem} (known also as Dini's
theorem, see for example \cite{Dini}) to the equation (\ref{quinticAppendix}) around $A = 0$
and to the equation (\ref{quinticv}) around $x = 0$. One can then prove that, to each of the roots
$v_0$, $v_{\pm}$ of (\ref{inner equation}), it is possible to associate a local solution of
(\ref{quintic}) def\mbox{}ined in a neighbourhood of $\r = 0^+$, which are respectively the diverging
inner solution $\textbf{D}$ and the f\mbox{}inite inner solutions $\textbf{F}_{\pm}$. Likewise, it can
be shown that it is possible to associate a local solution of (\ref{quintic}) def\mbox{}ined in a
neighbourhood of $\r = + \infty$, to each of the simple roots of (\ref{asymptotic equation}). These
solutions are the asymptotic solutions $\textbf{L}$, $\textbf{C}_{+}$, $\textbf{C}_{-}$, $\textbf{P}_{1}$ and $\textbf{P}_{2}$.
In the case $\textbf{P}_{1} = \textbf{P}_{2} \equiv \textbf{P}$, a separate analysis is needed. It can be shown
that for $\a > 0$ and $\b = c_{+}\, \aq$ there are no local solutions of (\ref{quintic})
which tend to $\textbf{P}$ when $\r \to + \infty$, and the same holds for $\a < 0$ and $\b =
c_{-}\, \aq$. On the other hand, for $\a > 0$ and $\b = c_{-}\, \aq$ there
are \emph{two} dif\mbox{}ferent local solutions of (\ref{quintic}) which tend to
$\textbf{P}$ when $\r \to + \infty$, and the same holds for $\a < 0$ and $\b = c_{+}\, \aq$.
Despite having the same limit for  $\r \to + \infty$, these two local solutions are dif\mbox{}ferent when $A \neq 0$:
we then call $\textbf{P}_{1}$ the solution which in absolute value is smaller, and $\textbf{P}_{2}$
the solution which in absolute value is bigger. Therefore, on the boundaries between the three-roots-at-inf\mbox{}inity
regions and the f\mbox{}ive-roots-at-inf\mbox{}inity regions, for $\a \gtrless 0$, $\b = c_{\pm}\, \aq$ there
are three asymptotic solutions of (\ref{quintic}), while for $\a \gtrless 0$, $\b =
c_{\mp}\, \aq$ there are f\mbox{}ive asymptotic solutions of (\ref{quintic}).

We now turn to the discussion of the leading behaviours of the inner and asymptotic solutions. For the f\mbox{}inite inner solutions $\textbf{F}_{\pm}$ and f\mbox{}inite non-zero asymptotic solutions $\textbf{C}_{\pm}$ and $\textbf{P}_{1,2}$, the behaviour is
\be
h(\r) = C + R(\r)
\ee
where $C \neq 0$ is their limiting value, and $R$ is respectively such that $\lim_{\r \to 0^+}
R = 0$ (inner solutions) and $\lim_{\r \to + \infty} R = 0$ (asymptotic solutions). For the
asymptotic decaying solution $\mathbf{L}$ and the inner diverging solution $\mathbf{D}$, a more
detailed study is worthwhile.
\\
\\
\small
\textbf{Asymptotic decaying solution $\mathbf{L}$}
\normalsize
\\
\\
Let's consider the solution $\mathbf{L}$, which satisf\mbox{}ies $\lim_{\r \to +\infty} h(\r) = 0$. Dividing
the quintic equation (\ref{quinticAppendix}) by $h$, we get
\be
\frac{3}{2} \, \bq \, \ha{4} - \big( \aq + 2 \b \big) \, \ha{2}
+ 3 \, \a \, h - \frac{3}{2}  = \bigg( \frac{\r_v}{\r} \bigg)^{\! 3} \, \bigg( \frac{1}{h} - 3 \, \b \, h \bigg)
\ee
The left hand side has a f\mbox{}inite limit when $\r \to +\infty$, so the same has to hold for the right hand
side: taking this limit in the equation above gives
\be
\lim_{\r \to +\infty} \bigg( \frac{\r_v}{\r} \bigg)^{\! 3} \, \frac{1}{h} = - \frac{3}{2}
\ee
which implies that
\be
h(\r) = - \frac{2}{3} \, \bigg( \frac{\r_v}{\r} \bigg)^{\! 3} + R(\r)
\ee
with $\lim_{\r \to +\infty} \r^3 R(\r) = 0$.
\\
\\
\small
\textbf{Inner diverging solution $\mathbf{D}$}
\normalsize
\\
\\
Let's consider now the solution $\mathbf{D}$, which satisf\mbox{}ies $\lim_{\r \to 0^+} \abs{h(\r)} = +\infty$.
Dividing the equation (\ref{quinticv}) by $v^3$, one f\mbox{}inds that
\be
\label{divergingv}
\va{2} - 3 \, \b = \bigg( \frac{\r}{\r_v} \bigg)^{\! 3} \, \frac{1}{\va{3}} \, \Big( - \frac{3}{2} \, \va{4}
+ 3 \, \a \, \va{3} - \big( \aq + 2 \b \big) \, \va{2} + \frac{3}{2} \, \bq \Big)
\ee
One more time, the left hand side has a f\mbox{}inite limit when $\r \to 0^+$, so the same should hold for the right hand side. Therefore, the $\rho \to 0^+$ limit in the equation above gives
\be
\lim_{\r \to 0^+} \bigg( \frac{\r}{\r_v} \bigg)^{\! 3} \, \frac{1}{\va{3}} = - \frac{2}{\b}
\ee
and so
\be
\label{vrest}
v(\r) = - \sqrt[3]{\frac{\b}{2}} \, \frac{\r}{\r_v} + \mathrm{R}(\r)
\ee
with $\lim_{\r \to 0^+} \mathrm{R}(\r)/\r = 0$.
To understand the behaviour of the gravitational potentials (\ref{solfab})-(\ref{solnab}) in this case, it is useful to calculate the next to leading order behaviour. In fact, it turns out that, after going back to $h = 1/ v$, the leading behaviour precisely cancels the Schwarzschild-like contribution, so to understand if the gravitational potentials are f\mbox{}inite at the origin it is essential to know how $\mathrm{R}$ behaves for very small radii. Inserting (\ref{vrest}) into (\ref{quinticv}) and dividing by $x^5$, one obtains taking the limit $\r \to 0^+$ that
\be
\lim_{\r \to 0^+} \frac{\mathrm{R}}{x^3} = \frac{1}{9 \, \b} \bigg( \aq + \frac{3}{2} \, \b \bigg)
\ee
where $x = \r / \r_v$. We have then
\be
v(\r) = - \sqrt[3]{\frac{\b}{2}} \, \frac{\r}{\r_v} + \mathcal{N} \, \Big( \frac{\r}{\r_v} \Big)^{\! 3} + \mathcal{R}(\r)
\ee
where
\be
\mathcal{N} = \frac{1}{9 \, \b} \, \Big( \aq + \frac{3}{2} \, \b \Big)
\ee
and $\lim_{\r \to 0^+} ( \mathcal{R}(\r) / \r^3 ) = 0$. Finally, going back to the function $h$ we get
\be
h(\r) =  - \sqrt[3]{\frac{2}{\b}} \, \frac{\r_v}{\r} - \mathcal{M} \, \frac{\r}{\r_v} + \mathscr{R}(\r)
\ee
where
\be
\mathcal{M} = \frac{1}{9} \, \sqrt[3]{\frac{4}{\b^5}} \, \bigg( \aq + \frac{3}{2} \, \b \bigg)
\ee
and $\lim_{\r \to 0^+} (\mathscr{R}(\r)/\r) = 0$. It can be shown that in the special case $\aq + 3 \, \b / 2 = 0$, the next to leading order term scales as $\r^2$ instead of $\r$, and that $\lim_{\r \to 0^+} (\mathscr{R}(\r)/\r^2) = 0$.

Therefore, we can conclude that in general the diverging inner solution $\textbf{D}$ is such that
\be
h(\r) =  - \sqrt[3]{\frac{2}{\b}} \, \frac{\r_v}{\r} + R(\r)
\ee
where $\lim_{\r \to 0^+} (R(\r)/\r)$ is f\mbox{}inite.
%In the special case $\aq + 3 \, \b / 2 = 0$ one gets
%\be
%h(\r) = \sqrt[3]{\frac{3}{\aq}} \, \frac{\r_v}{\r} - \frac{1}{2 \a} \, \sqrt[3]{\frac{3}{\aq}} \, \Big( \frac{\r}{\r_v} \Big)^{\! 2} +  \mathscr{R}(\r)
%\ee
%with $\lim_{\r \to 0^+} (\mathscr{R}(\r)/\r^2) = 0$.

\end{appendix}

\end{document}